\documentclass[aps,prl,twocolumn]{revtex4}

\usepackage{graphicx}
\usepackage{dcolumn}
\usepackage{amsmath}
\usepackage{amssymb}
\usepackage{wasysym}
\usepackage{color}

\usepackage{mathtools}
\usepackage{booktabs}

\begin{document}

\title{Magnetic-Field-Tuned Critical End Point and High-Field Phases in HoMn$_6$Sn$_6$}

\author{Joanna Blawat$^1$, John Singleton$^1$, Junho Seo$^2$, Charles Mielke III$^2$, Tyler J. Slade$^{3,4}$, 
Paul~C.~Canfield$^{3,4}$, Philip J. W. Moll$^2$, Ross D. McDonald$^1$, and Robert J. McQueeney$^{3,4}$
}

\affiliation{
$^1$National High Magnetic Field Laboratory Pulsed-Field Facility, Los Alamos National Laboratory, 
Los Alamos, New Mexico 87545, USA\\
$^2$Max Planck Institute for the Structure and Dynamics of Matter, Hamburg 22761, Germany\\
$^3$Ames National Laboratory, U.S. DOE, Iowa State University, Ames, Iowa 50011, USA\\
$^4$Department of Physics and Astronomy, Iowa State University, Ames, Iowa 50011, USA}
\date{\today}

\begin{abstract}\noindent
Topological Kagome $R$Mn$_6$Sn$_6$ materials, where $R$ is a rare-earth, exhibit a rich variety of magnetic phases 
which provide an opportunity to understand  how different spin textures affect topological and transport properties through 
the application of magnetic field $H$. In this work, we report experiments on single-crystal HoMn$_6$Sn$_6$, made possible 
by pulsed magnetic fields of up to 60~T, that permit high-field features of the field-temperature phase diagram to be studied 
in detail using magnetometry. Complementary, sensitive transport experiments over the same field range employ 
focused-ion-beam fabrication of novel strain-relieved HoMn$_6$Sn$_6$ devices. In agreement with mean-field calculations, we find 
a metamagnetic transition at 45~T with $H || c$ between canted phases that ends in a critical end point at about 90~K.
 As far as we are aware, this is the only critical end-point that is observed within the RMn$_6$Sn$_6$ kagome metals, and its temperature 
 and field far exceed those of analogous phenomena in other materials. 
 In the low-temperature metamagnetic region, HoMn$_6$Sn$_6$ shows large magnetoresistance, reaching 900~\% at 60~T, 
 in concert with Shubnikov de Haas oscillations that suggest a slight Fermi-surface 
contraction as the metamagnetic transition is crossed. 
A phase analysis of the oscillations tentatively suggests non-trivial band topology, although a firmer attribution must await detailed bandstructure calculations. Near the critical end point, we find deviations 
from Fermi-liquid behavior in the temperature-dependent magnetoresistance. The Hall resistivity indicates that the dominant carriers 
are holes.  In the metamagnetic region, the anomalous Hall contribution is not proportional to magnetization, and peaks at about 
40~K, suggesting the role of enhanced spin fluctuations near the critical point. At higher temperatures, the anomalous Hall effect reappears at low fields, showing that the magnetization (and its fluctuations) 
provided by the Ho and Mn sublattices independently affect the electrical conductivity in distinct ways.
\end{abstract}

\maketitle 

\section{Introduction}
Since the discovery of topological states in semimetals~\cite{r1}, the search for new topological materials has been extremely 
active due to their potential application in areas such as spintronics and quantum computing~\cite{r2,r3,r4}. 
As well as exploring the fundamental connection between (broken) crystallographic symmetries and topological protection, 
the interplay between topology, magnetism, superconductivity, and strong correlations has garnered special attention because of the 
promise of new and exotic quantum phenomena in topological matter~\cite{r5,r6,r7,r8}. In addition to band crossing originating 
from the crystal symmetry, compounds with strong spin-orbit coupling can exhibit topologically protected band crossing, 
giving rise to a range of novel topological phases~\cite{r9,r10}. The number of bands crossing each other (degeneracy) depends 
on the crystal symmetry and magnetism. For example, bulk Dirac band crossing can occur when both the 
inversion $\cal{P}$ and time-reversal $\cal{T}$ symmetries are preserved, whereas in ferromagnetic and more complex 
magnetic structures with broken $\cal{T}$ symmetry, degeneracy of the Dirac bands is further lifted, potentially resulting in 
pairs of Weyl nodes with opposite chirality~\cite{r1,r11}. 

An attractive platform for studying this intersection of topology, strong correlations and magnetism is provided by the 
$R$Mn$_6$Sn$_6$ ($R$166) family, where $R$ is a rare earth; these kagome metals exhibit complex magnetic orders 
and spin textures\cite{r12}. Importantly, their electronic structure embodies all of the relevant features: (1)~flat bands -- responsible 
for strong electron correlations, (2)~Dirac cones -- bringing non-trivial topology, and (3)~a van-Hove singularity which causes instabilities
in the system, resulting in long-range magnetic order~\cite{r6,r7,r13,r14,r15,r16,r17,r18,r19,r20}. 
The $R$166 materials contain Mn kagome bilayers, with strong ferromagnetic intralayer and interlayer interactions, 
that are separated by rare-earth triangular layers. The complex magnetism observed across the family, in some cases involving 
spontaneous reorientation of colinear ferrimagnetic structures~\cite{r21}, arises from competing interactions between 
Mn and R ions. The magnetic ground state within the $R166$ family changes from ferrimagnetic ($R$~=~Tb, Gd), to canted 
ferrimagnetic ($R$~=~Ho and Dy), to antiferromagnetic helical order ($R$~=~Y, Tm, Lu)~\cite{r6}, 
with $R$~=~Er being ferrimagnetic at low temperatures but changing to helical order at high temperatures~\cite{r6}. 
The magnetic structure of TbMn$_6$Sn$_6$, where Tb spins are oriented antiparallel to Mn spins, supports Chern gapped 
Dirac fermions~\cite{r14}. As $R$ changes from Gd to Tm, the exchange coupling between $R$ $4f$ electrons and Mn $3d$ electrons 
weakens, and the Chern gap becomes smaller ~\cite{r6}.

The above-mentioned magnetic phases and topological properties are clearly manifested in magnetotransport; 
$R$~=~Tb, Dy, Ho, Gd and Y exhibit both topological and anomalous Hall effects~\cite{r21,r22,r23}, with, in some cases, the 
differing spin dynamics of the $R $ and Mn sublattices playing an important role~\cite{r23}.
The $R$166 system is thus a very flexible tool for studying the interplay between magnetism and topology, with the 
magnetism and non-trivial topological phases being tuned and controlled via the rare-earth element $R$ sitting between the kagome planes~\cite{r6,r21,r24,r25}.

Recent mean-field calculations~\cite{r26} predict that $R$166 systems exhibit an even richer variety of magnetic phases 
under external magnetic field $H$. The field provides another tuning parameter to scan through different spin textures, 
this time within a single compound, potentially allowing access to different topological phases affected by magnetic symmetry~\cite{r26}. 
For example, HoMn$_6$Sn$_6$ is predicted~\cite{r26} to exhibit a critical end-point for $H || c$ at just over 40~T and 50~K.  
To our knowledge, such an end-point has not previously been observed at such high temperatures and magnetic fields; 
therefore, its influence on physical properties, such as electrical conductivity and Hall effect has not been studied. 

Thus motivated, we have performed magnetization and magnetotransport measurements on HoMn$_6$Sn$_6$ in pulsed 
magnetic fields of up to 60~T and temperatures down to 600~mK. The magnetization data confirm the presence of a first-order 
metamagnetic transition terminating at a critical end-point at 45~T and 90~K. Using novel HoMn$_6$Sn$_6$ devices fabricated by 
focused-ion-beam (FIB) techniques, designed both to enhance the voltages corresponding to the resistivity tensor components of interest 
and to relieve strain due to thermal contraction, we were able to study the transport behavior in pulsed magnetic fields of up to 60~T. 
This enables measurements in both the metamagnetic and critical endpoint regions of the phase diagram, revealing large magnetoresistance 
and quantum oscillations at low temperatures, anomalous contributions to the Hall conductivity, and the role of fluctuations. 
Analysis of the quantum oscillations indicates a change in Fermi surface at the metamagnetic transition, whilst non-trivial band topology 
is maintained. 

\section{Experimental Methods}
Single crystals of HoMn$_6$Sn$_6$ were grown from excess 
Sn using the flux growth method. 
A nominal (HoMn$_6$)$_5$ to Sn$_{95}$ molar ratio of elemental Ho (Ames Laboratory, $> 99.9~\%$), 
Mn (Research Organic/Inorganic Chemical Corp, 99.995~\%), 
and Sn (Alfa Aesar, 99.99~\%) was weighed and 
loaded into the growth side of a 2~mL Canfield 
Crucible Set (CCS)~\cite{growth1,growth2}. 
The crucibles were then flame-sealed under 
vacuum inside a fused silica ampoule with a small 
amount of silica wool placed above and below the 
crucible set for cushioning. 
The crucibles were heated to $1180^\circ$C, held at temperature for 6~hours, then cooled slowly to 
$575^\circ$C over approximately 300~hours. 
At the final temperature, the ampoules were removed from the furnace and the excess flux was decanted using a centrifuge with metal rotors and cups. 
After cooling, the crucibles were opened to reveal plate-like, hexagonal crystals. The crystals were confirmed to be HoMn$_6$Sn$_6$ using energy dispersive X-ray spectroscopy (EDS) and powder X-ray diffraction (PXRD).

\begin{figure}
\begin{center}
\includegraphics[width=1.0\linewidth]{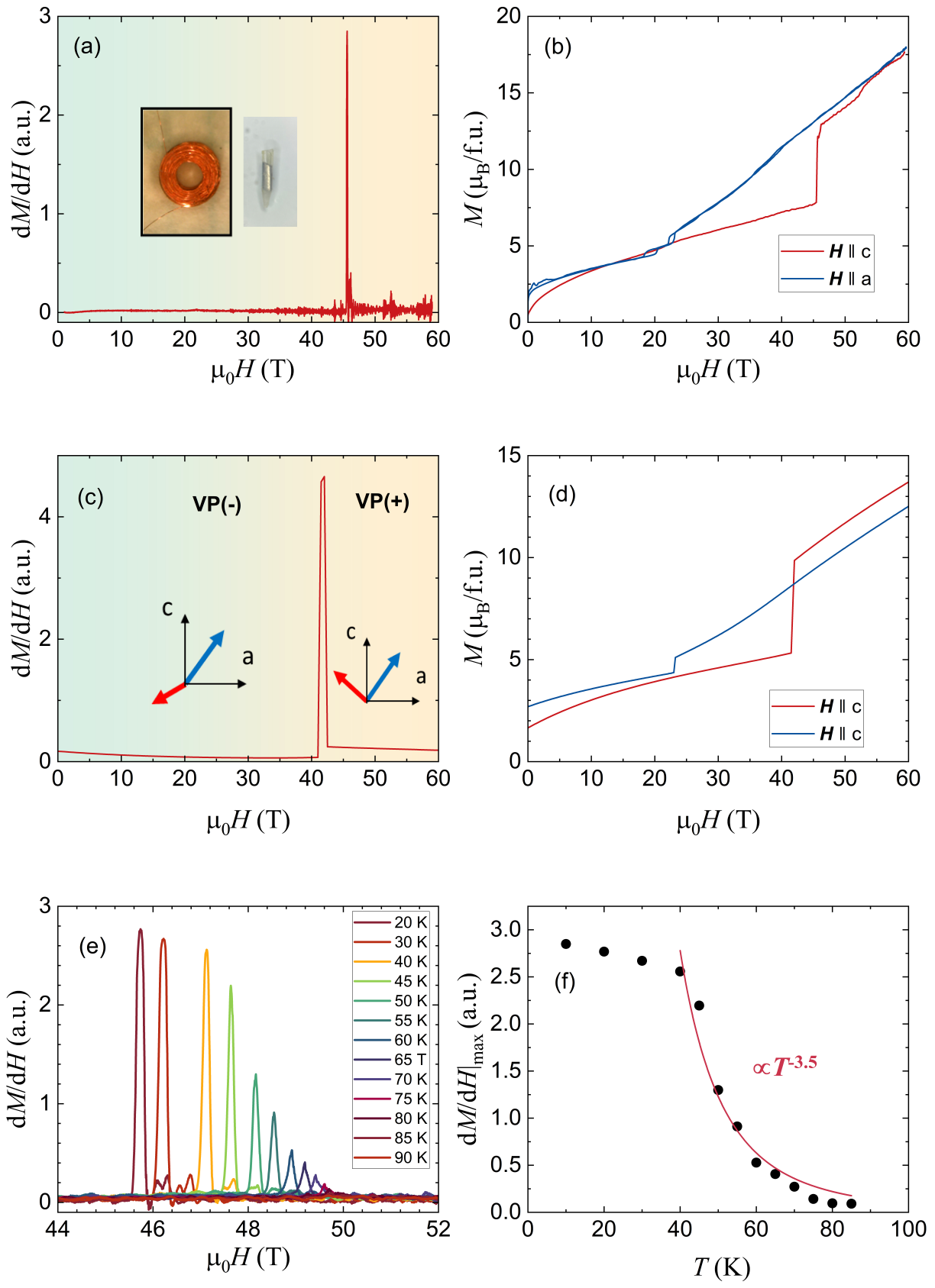}
\vspace{-4mm}
\caption{(a)~Magnetic susceptibility ${\rm d}M/\rm{d}H$ versus magnetic field $H$, recorded at
a temperature $T=0.6$~K. The inset showes the 1500-turn, 1.5~mm bore coil used for the measurements (left) and the sample in its
1.3~mm outer diameter ampoule (right).
(b)~Magnetization $M$ versus $H$ for $H||a$ and $H||c$, recorded at 0.6~K.
(c)~and(d) Mean-field calculations of the corresponding experimental data in panels (a) and (b) respectively, using the mean-field model summarized in the Appendix~\cite{r26}. The arrows shown 
in (c) represent the orientations of the Ho (red) and Mn (blue) moments suggested by the mean-field model. 
The labels refer to phases tabulated in Table~\ref{tbl:structures}. 
(e)~Magnetic susceptibility versus $H$ for several different
temperatures $T$ listed in the inset key. Note that the peaks shift upwards with increasing $T$ and become smaller.
(f)~Heights of susceptibility peaks plotted as a function
of $T$ (points). The red curve is $\propto T^{-3.5}$. 
}
\vspace{-9mm}
\label{fig1}
\end{center}
\end{figure}

The high field experiments were performed 
at the National High Magnetic Field Laboratory Pulsed-Field Facility in Los Alamos. 
Magnetization was measured using the compensated-coil extraction magnetometer~\cite{r27}; an oriented piece of crystal is
placed in an ampoule (1.3~mm outer diameter, 1~mm bore)which can be moved in and out of the coil (see Fig.~\ref{fig1}~(a)). The voltage induced in the coil by 
the time-dependent (pulsed) field $H(t)$ is proportional to $({\rm d}M/{\rm d}t)$. To obtain the magnetization 
$M(H)$, data recorded with the sample in the coil are integrated with respect to time and then similarly integrated data 
recorded with the coil empty are subtracted. The differential susceptibility is obtained by dividing $({\rm d}M/{\rm d}t)$ by 
$({\rm d}H/{\rm d}t)$, the latter measured using a separate coil also located at the field center~\cite{r27}.

The resistivity and Hall effect data were measured on microstructured devices fabricated using Focused Ion Beam (FIB) techniques. 
First, a lamella was cut off from a bulk piece of single crystal and transferred {\it in-situ} via micromanipulator to a gold coated 
SiN$_x$ membrane substrate. The sample was welded via Pt-deposition to the substrate, and the meander shaped springs were 
fabricated in the membrane. Similar procedures are described in detail in ref.~\cite{r39}, but here the HoMn$_6$Sn$_6$ was fixed 
on one side to counteract the impulsive torque which might occur when measuring magnetic samples in pulsed fields. 
To minimize the noise due to relative motion of wires during the pulsed-field measurements, $50-$gauge insulated copper wires 
were attached to each Pt contact and then twisted in current $(I)$, resistance $(R_{xx})$ and Hall resistance $(R_{xy})$ pairs. 
The measurement was performed using a $500~\mu$A AC current, with a typical frequency of $f\approx 65~$kHz.

\section{Experimental Results} 
\subsection{Magnetic phase diagram}
The magnetic susceptibility, $({\rm d}M/{\rm d}H)$, was measured as a function of temperature at fields of up to 60 T; 
a typical curve with $H || c$ is shown in Fig.~\ref{fig1}~(a), revealing a large peak at 45~T. Such peaks correspond to rapid changes in 
magnetization and are used to locate the positions of phase transitions. Magnetization $M(H)$ data are plotted in Fig. 1(b). 
With $H || a$ at 600~mK, the dominant feature is a transition at 20~T subdivided into three step-like features that result in three 
peaks in the corresponding susceptibility data. For $H || c$, there is a sharp, step-like 
metamagnetic transition in $M(H)$ close to 45~T, associated with the peak in
the susceptibility at the same field (Fig.~\ref{fig1}~(a)); as discussed below,
we identify this feature with the phase boundary associated with the critical end-point

\begin{table}[htbp]
\caption{Magnetic structures in $\text{HoMn}_6\text{Sn}_6$ with $H \parallel z \equiv c$.}
\begin{tabular}{|c|c|}
\hline\hline
Phase & Description  \\
\hline 
VP(-) & Vertical-plane canted,\\ 
(low $T$, low $H$) & Ho moment proj. antiparallel to $H$  \\
\hline
VP(+) &  Vertical-plane canted, \\
low $T$, high $H$ & Ho moment proj. parallel to $H$ \\
\hline
FIM$a$ & Ferrimagnetic with moments along $a$  \\
(high $T$, $H=0$) & ~\\
\hline
FIM$c$ & Ferrimagnetic with moments along $c$  \\
(high $T$, low $H$) & ~\\
\hline
Fully polarized & All moments along $c$  \\
(high $T$, high $H$) & ~\\
\hline\hline
\end{tabular}
\label{tbl:structures}
\end{table}
To help understand the reasons for the critical endpoint, it is useful to summarize the phases predicted by 
mean-field theory~\cite{r26}, which are tabulated in Table~\ref{tbl:structures};
for convenience, the Hamiltonian and model parameters are described briefly in the Appendix.
For $H=0$, the interactions in HoMn$_6$Sn$_6$ favor antiparallel Ho and Mn moments, lying tilted in the ac-plane ~(see Appendix and Ref.~\cite{r26}). 
As the Mn sublattice has a total moment of $13\mu_{\rm B}$ 
and the Ho sublattice $10\mu_{\rm B}$, the application of finite $H || c$  initially produces a tendency for the
the Mn moments to rotate towards $H$ and the Ho moments away from it. In other words, at $H=0$ the $R$ and Mn moments are collinear and tilted away from the $c-$axis 
within the $ac-$plane. Upon application of a magnetic field, the Mn moments gradually rotate toward the field direction, while the $R$ moments remain pinned along their easy direction because of the large magnetocrystalline anisotropy (MAE) energy. At the critical field, the $R$ moments abruptly switch from the MAE minimum in the lower half of the $ac-$plane to the equivalent minimum in the upper half. At low $T$, this results
in the moments forming a canted arrangement in the $ac-$plane (Vertical-plane canted or VP).  
The metamagnetic transition in $M(H)$ close to 45~T represents a change 
from the Ho moments making an angle $>90^\circ$ with $H$
(VP(-) phase) to them making an angle $<90^\circ$ (VP(+) phase). At low $T$, this transition is step-like,
resulting in a peak in ${\rm d}M/{\rm d}H)$ [see Fig.~\ref{fig1}(a-e)]; however, the two VP phases have the same symmetry, 
such that a critical end point is obtained owing to quenching of the Ho magnetic anisotropy barrier with increasing $T$.
Beyond the critical point, free rotation of the Ho moment is associated with crossover behavior~\cite{r26}.

\begin{figure}
\begin{center}
\includegraphics[width=0.88\linewidth]{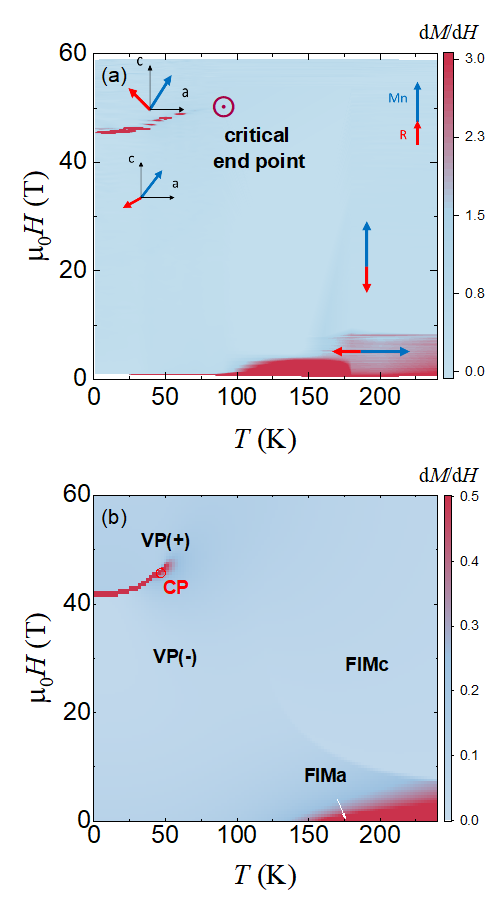}
\vspace{-4mm}
\caption{(a)~Experimental susceptibility contour plot as a function of $T$ and $H$ for $H||c$.
The phase boundary terminating in the critical endpoint is clearly seen in the top left-hand corner.
(b)~Analogous theoretical contour plot predicted by the mean-field model (Appendix A~\cite{r26}).
In (a), the arrows represent the orientations of the Ho (red) and Mn (blue) moments suggested by the mean-field model;
in (b), the labels refer to phases tabulated in Table~\ref{tbl:structures}.  
}
\vspace{-9mm}
\label{fig2}
\end{center}
\end{figure}

To build a comprehensive $H-T$ phase diagram with $H || c$ that confirms the above assertions, 
${\rm d}M/{\rm d}H$ data were taken at many temperatures 
between 600~mK and 240~K.  Figure~\ref{fig1}(e) shows the peak in $({\rm d}M/{\rm d}H)$ close to 45~T; the peak
becomes gradually smaller with increasing temperature, disappearing completely at about 90~K. 
This behavior is consistent with a phase boundary termination at a critical end point.
The resulting experimental phase diagram is presented in Fig.~\ref{fig2}(a); 
the equivalent theoretical phase diagram produced by the model (Appendix and Ref.~\cite{r26}) is shown in Fig.~\ref{fig2}(b).
At just under 50~T, the critical field at the end-point is in good agreement with mean-field calculations; however, the 
experimental endpoint temperature of about 90~K is significantly higher than the predicted value of 50~K. The variation of 
the critical field with temperature is also less steep in the experimental data than it is in the model. 

Fluctuations are expected to be associated with the approach to the critical endpoint; Fig.~\ref{fig1}(f) shows the behavior of 
the maximum value of $({\rm d}M/{\rm d}H)$ along the phase boundary as the endpoint approaches. 
Consistent with thermal quenching of the Ho anisotropy barrier between VP(-) and VP(+) phases, 
the maximum in $({\rm d}M/{\rm d}H)$ at first decreases slowly with rising $T$, and then more rapidly
above 40~K, vanishing at the end point; in this higher temperature region, its height scales approximately as $T^{-3.5}$ (Fig.~\ref{fig1}(f)). 
For symmetry-breaking magnetic phase boundaries that extend 
from $T=0$ at finite $H$ to $H=0$ at finite $T$, a qualitatively similar dependence of the susceptibility peak
$\propto T^{-m}$ has been taken as an indication of quantum fluctuations near to a critical point~\cite{r31,r32,r33} . 
However, this tendency has only been reported for low $T$ and low-field $2^{\rm nd}$ order phase transitions; 
in our much higher $T$ data, the unusual temperature exponent may rather be associated with dynamical spin 
and/or thermal fluctuations analogous to those reported in TbMn$_6$Sn$_6$~\cite{r21,r24} 
and CeOs$_4$Sb$_{12}$~\cite{r34,r35}.

Finally, we note that, consistent with previous data obtained at 4.2~K and with fields of up to 55~T~\cite{r20}, 
our $M(H)$ data do not saturate for field of up to 60~T; when compared with theoretical predictions~\cite{r26}, 
this suggests that other metamagnetic transitions might occur at higher fields. In fact, further increasing the field continuously rotates both the Mn and $R$ moments toward the field direction. The system therefore evolves into the fully polarized phase through a second-order transition, which is predicted to occur only at very high fields ($>150$~T)~\cite{r26}.

\begin{figure*}
\begin{center}
\includegraphics[width=0.80\linewidth]{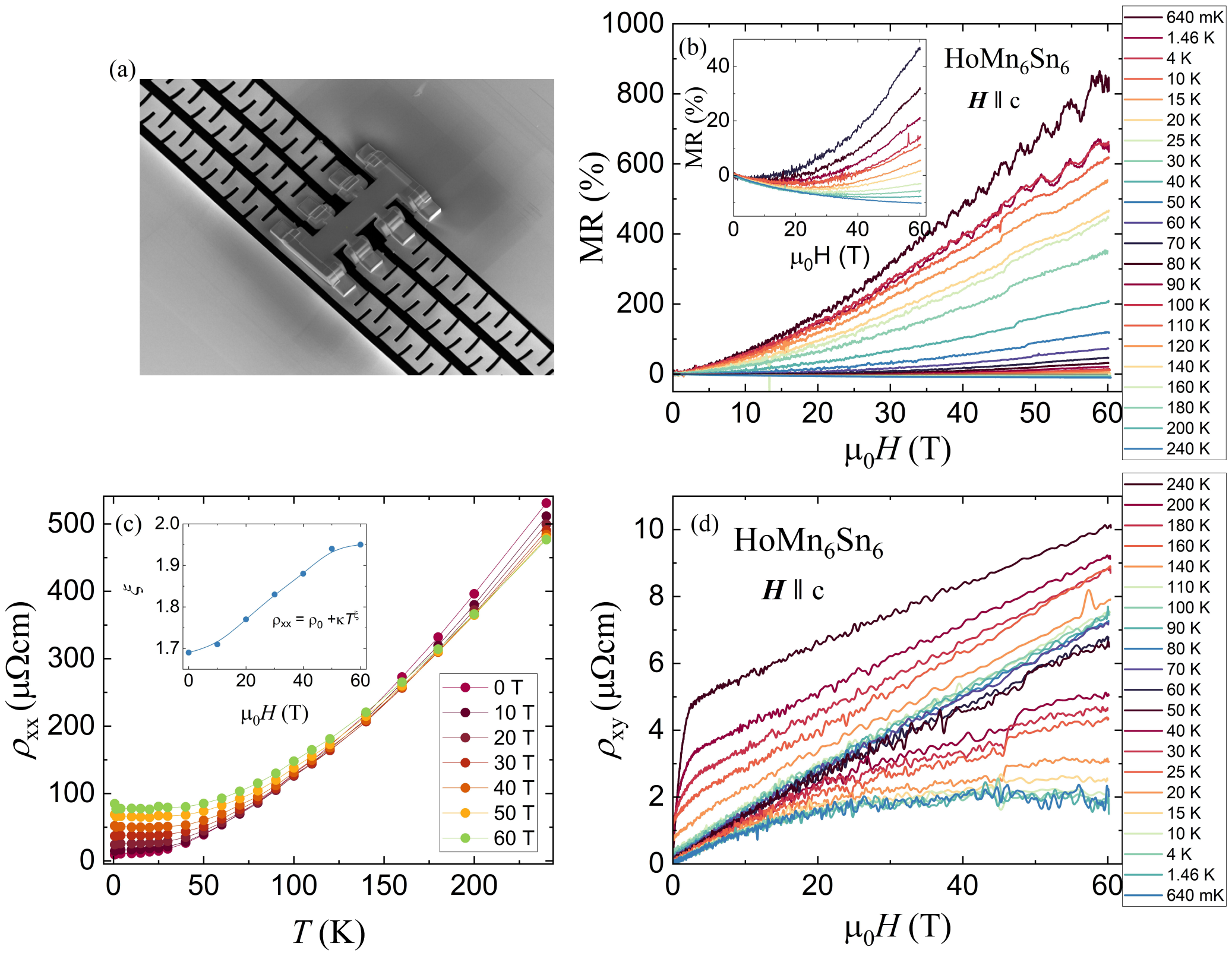}
\vspace{-4mm}
\caption{(a)~A strain-relieved device fabricated by FIB and used in the magnetoresistance and Hall-effect experiments.
(b) Magnetoresistance (MR) data plotted as a function of $H$ for several different temperatures and $H||c$.
Shubnikov-de Haas oscillations are clearly seen for lower values of $T$.
The inset shows the high-temperature field sweeps om an expanded vertical scale, so that the
emerging negative magnetoresistance is more easily visible.(c)~Values of $\rho_{xx}$ for
different magnetic fields plotted versus $T$; the points are extracted from constant-$T$
swept-field data such as those in (b).
(d)~Hall resistivity $\rho_{xy}$ versus $H$ for several different temperatures.
}
\vspace{-9mm}
\label{fig3}
\end{center}
\end{figure*}

\subsection{Magnetotransport}
Low-field magnetotransport measurements have been reported for HoMn$_6$Sn$_6$ and several other 
$R$Mn$_6$Sn$_6$ compounds~\cite{r21,r24,r25,r36}, materials that are generally good metals with relatively 
high carrier concentrations. Such experiments often show a large intrinsic anomalous Hall effect (AHE), along 
with unsaturated positive magnetoresistance (MR) at low temperatures that evolves to negative MR
as $T$ increases~\cite{r22,r23,r35}. These and other transport phenomena are sensitive to details of the underlying 
ferrimagnetic order~\cite{r22}  and its spin dynamics~\cite{r24}, with the development of Berry curvature hosted by 
certain magnetic phases~\cite{r21,r22} providing further interest. In view of this connection between microscopic details 
of the magnetic ground state and features in the electrical conductivity, it is informative to study the latter close to the 
high-field metamagnetic transition and the region close to the critical endpoint where fluctuations in the magnetization are large. 
In this section, we therefore focus on magnetotransport properties just below the metamagnetic transition $(T\leq 70~$K), 
in the vicinity of the critical point ($70\leq T\leq 100$~K), and beyond ($T \geq 100~$K). 

\begin{figure*}
\begin{center}
\includegraphics[width=0.99\linewidth]{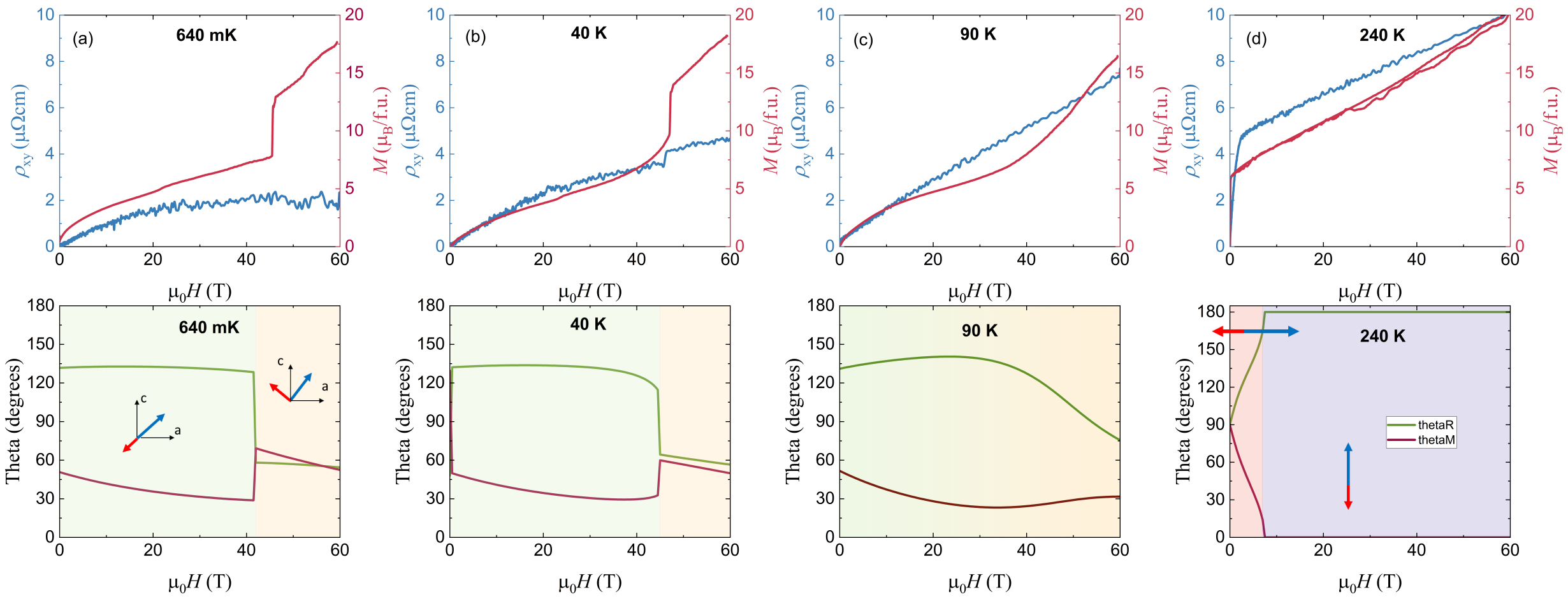}
\vspace{-4mm}
\caption{Examples of $\rho_{xy}$ plotted alongside the corresponding $M(H,T)$ data and
mean-field model predictions of the orientations of the Mn and Ho moments for four different temperatures; 
$\theta$ is the angle between the Mn and Ho moments and the $c-$axis.
The temperatures are 640~mK~(a),
40~K~(b), 90~K~(c) and 240~K~(d).
At 640~mK~(a), the lack of any visible change in $\rho_{xy}$ around 45~T, despite the large step in $M(H)$ due to the metamagnetic transition, suggests that the latter does not result in a very significant change in the electronic bandstructure. 
}
\vspace{-9mm}
\label{fig4}
\end{center}
\end{figure*}

\begin{figure}
\begin{center}
\includegraphics[width=0.7\linewidth]{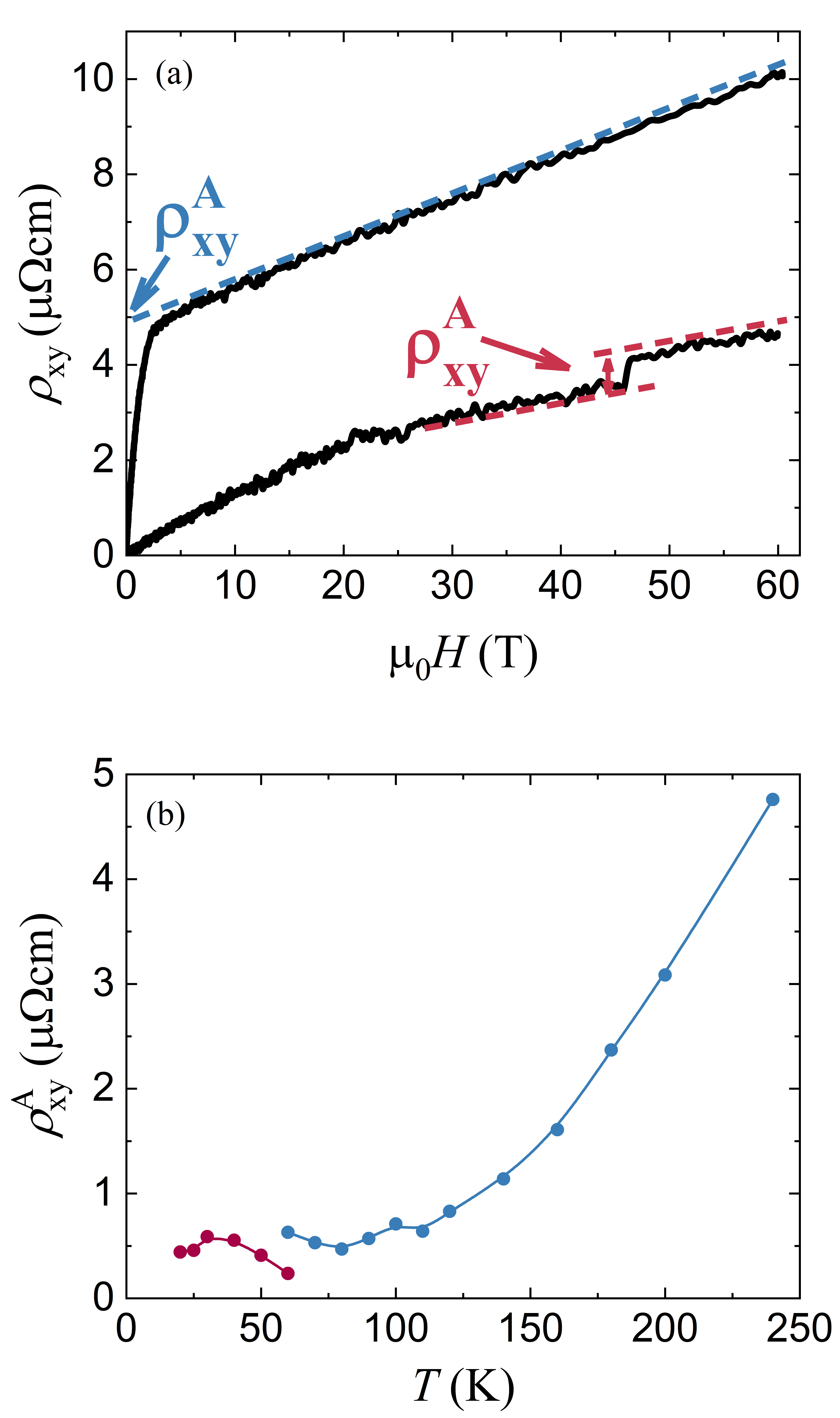}
\vspace{+4mm}
\caption{(a)~Procedure for deriving $\rho^{\rm A}_{xy}$ from high-temperature (blue) and 
low-temperature (red) $\rho_{xy}$ data. For high-temperatures, the quasi-linear high-field
data are extrapolated back to $H=0$; the intercept gives the component due to the 
rapid increase in $M$ at low fields. At lower temperatures, where the metamagnetic
transition is visible as a step in the Hall data, $\rho^{\rm A}_{xy}$ is taken to be the difference between
$\rho_{xy}$ above and below the transition.
(b)~Extracted anomalous Hall components; red points are associated with Ho
fluctuations close to the metamagnetic transition; blue correspond to rotation of the Mn moment from planar to uniaxial.
}
\vspace{-9mm}
\label{fig5}
\end{center}
\end{figure}

However, during initial experiments we found that samples of HoMn$_6$Sn$_6$ are extremely sensitive to strain 
and thermal contraction, such that the standard procedure of cutting a Hall bar using FIB
lithography on hand-polished samples did not yield thermally stable devices. (Such sample processing is necessary 
to give good signal-to-noise ratios in sensitive $\rho_{xx}$ and $\rho_{xy}$ measurements of this high-conductivity metal 
in the extreme electromagnetic environment of pulsed magnetic fields.) To mitigate these challenges, we prepared 
strain-relieved devices by FIB microstructuring both the HoMn$_6$Sn$_6$ sample and the supporting SiN$_x$ membrane, 
as illustrated in Fig.~\ref{fig3}(a). The device dimensions are ${\rm 57~\mu m \times 12~\mu m \times 3.5~\mu m}$. 
Consequently, the data shown in Figs.~\ref{fig3} to \ref{fig5} represent an important experimental advance, resulting from magnetotransport measurements on strain-relieved devices in pulsed magnetic fields and sub-Kelvin temperatures.

Figure~\ref{fig3} shows magnetoresistance (MR), $\rho_{xx}$ and $\rho_{xy}$ for $H || c$ in all temperature regions.
Here, the magnetoresistance is defined as
\begin{equation}
{\rm MR(\%)} = 100\times \frac{(\rho_{xx}(H) - \rho_{xx}(0))}{\rho_{xx}(0)}, 
\end{equation}  
where $\rho_{xx}(H)$ and $\rho_{xx}(0)$ are the diagonal resistivity component $\rho_{xx}$ at finite field $H$ and at $H=0$
respectively; representative data 
 are shown in Fig.~\ref{fig3}(a). For $H||c$, the MR is positive and increases as ${\rm MR} \propto H^{1.5}$
at 640~mK, reaching 900~\% at 60~T with no sign of saturation. For $\mu_0H \geq 30~$T, Shubnikov de Haas (SdH) oscillations 
are clearly observed; their analysis is described in more detail below. At higher temperatures $(20~{\rm K}\leq T \leq 90~{\rm K})$,
a pronounced step in resistance is observed at the metamagnetic transition (close to 45~T), which fades away
as the critical endpoint is approached. Above the metamagnetic field, ${\rm MR}(H)$ increases approximately 
linearly with $H$. At 70 K, the MR changes sign and becomes negative at low $H$. 
By 140~K, it is negative over the whole field range, and reaches -15~\% at 240~K and 60~T (see Fig.~\ref{fig3}(a), inset).  The latter high-field trends of our data 
mirror the behavior seen in some earlier low-field experiments~\cite{r23}; this is not unexpected, as most compounds in the $R$166 family exhibit ferrimagnetic 
ordering (FIMc - see Table~\ref{tbl:structures}) in this field and temperature region ({\it i.e.,} above 100~K). 
 According to the mean-field calculations (see Appendix~\cite{r26}),  in the high-temperature, high field FIMc phase, the Ho moment is antiparallel to $H$ and will decrease with increasing field 
 ({\it i.e.,} Ho thermal fluctuations will increase) until $\approx 70~$T, above which the Ho moment becomes parallel to $H$ 
 and grows in size. On the other hand, the Mn thermal fluctuations are suppressed as its moment is parallel to $H$. 
 This suggests that the negative MR observed in the $R$166 family is likely the result of suppression of thermal fluctuations of the Mn moments,
 leading to a reduction of spin-disorder scattering.

Figure~\ref{fig3}(c)  shows $\rho_{xx}$ versus $T$ at constant $H$ extracted from the isothermal pulsed-field sweeps. 
As expected, $\rho_{xx}$ shows metallic ({\it i.e.,} positive $({\rm d}\rho_{xx}/{\rm d}T)$) behavior over the whole range of $H$ and $T$. 
To examine the evolution of this $T$-dependence with $H$, $\rho_{xx}$ data were fitted to
\begin{equation}
\rho_{xx} = \rho_0 + \kappa T^\xi,
\label{eqTfit}
\end{equation}
where $\rho_o$ and $\xi$ are fit parameters; derived $\xi$ values are plotted versus $H$ in Fig.~\ref{fig3}. 
For $H=0$, the $T$-dependence of the data is well captured by $\xi = 1.69$. 
However, as $H$ increases, $\xi$ grows smoothly, reaching $\xi = 1.96$ ({\it i.e.,} very close to the Fermi-liquid value
$\xi = 2$~\cite{r37}) at 50~T, just above the metamagnetic phase boundary. As described above (Table~\ref{tbl:structures}), the 
mean-field calculations show that the Ho moment goes from canted antiparallel to the field (VP(-)) to parallel (VP(+)).  
Taken in conjunction with the $M(H)$ data (Fig.~\ref{fig1}), this may suggest that the deviation from Fermi-liquid 
behavior below 45~T is due to Ho magnetic fluctuations, whilst above 45~T the Ho fluctuations are suppressed,
leading to the emergence of Fermi-liquid behavior. In previous low-field studies of TbMn$_6$Sn$_6$, magnetic fluctuations 
were found to influence transport properties~\cite{r7}.

Figure~\ref{fig3}(d) summarizes the Hall data; $\rho_{xy}$ stays positive up to 60~T for all $T$, suggesting that holes are the dominant 
charge carriers. At 640~mK, $\rho_{xy}$ exhibits curvature, usually attributed to multiband behavior~\cite{r37}.
Note that any change in $\rho_{xy}$ at around 45~T is rather small at this temperature, despite the dramatic step seen in magnetization [Fig.~\ref{fig1}(a,c)]; this suggests that the metamagnetic transition does not have a very significant effect on the electronic bandstructure (we shall return to this point below). 
The curvature of $\rho_{xy}$ becomes less pronounced as $T$ rises; between 10~K and 80~K, a clear step appears at the 
metamagnetic transition near 45~T, followed by linear Hall resistivity. Above 70~K ({\it i.e.,} the $T$ range approaching 
the critical endpoint) the high-field behavior of $\rho_{xy}$ is almost linear in $H$ up to 60~T.
Sudden changes in $\rho_{xy}$ as a function of $H$ are observed in two regions: (i)~at high fields $(>45~$T) 
and low temperatures $(<90~$K) corresponding to the metamagnetic transition between canted states; 
and (ii)~low fields $(<5~$T) and high temperatures corresponding to field-driven spin reorientation transition 
between collinear states.

To help visualize how these trends relate to the underlying magnetic configuration, Fig.~\ref{fig4} shows
four examples of $\rho_{xy}$ measured at different temperatures, plotted alongside the corresponding $M(H,T)$ data and
mean-field model predictions of the orientations of the Mn and Ho moments under the same conditions. 
We consider two contributions to $\rho_{xy}$:
\begin{equation}
\rho_{xy} = \rho_0(H) + \rho^{\rm A}_{xy}.
\end{equation}
Here $\rho_0(H)$ represents the ordinary (or conventional) Hall effect, and $\rho^{\rm A}_{xy} = 4\pi R_{\rm S} M_z$ is the 
anomalous Hall effect~\cite{r23}. 

In Figs.~\ref{fig4}(a) and (c), $\rho_{xy}$ varies smoothly with $H$, despite
the large step in $M(H)$ at 640~mK~(a) and the complex behavior of $M(H)$ at 90~K~(c). This observation suggests that
the anomalous Hall effect makes only a small contribution to $\rho_{xy}$ at these temperatures. 
On the other hand, at intermediate temperatures, represented by 40~K in Fig.~\ref{fig4}(b) [see also Fig.~\ref{fig3}(d)], the metamagnetic transition in $M(H)$ is mirrored by 
a small step in $\rho_{xy}$. As discussed above, the metamagnetic transition at around 45~T is attributed to the $z-$component of the Ho moment jumping 
parallel to $H$; in response, the Mn moment rotates away from $H$ (towards the $ab$ plane) to minimize the exchange energy.  
At low temperatures, there is little or no anomalous Hall response. However, the response grows as the temperature increases to about $30-40$~K,
before fading away as the critical endpoint is approached, suggesting a Ho-fluctuation-driven mechanism [see also the above discussion of Fig.~\ref{fig3}(c)]. 

At higher temperatures [Fig.~\ref{fig4}(d)], the magnetization of 
HoMn$_6$Sb$_6$ evolves from a planar ferrimagnet at low fields to one that is uniaxial and parallel to $H$ above a few Tesla~\cite{r26} (see Table~\ref{tbl:structures}). 
As previously reported for $T>120 ~$K~\cite{r23}, $\rho_{xy}$ increases significantly owing to a strong anomalous-Hall-effect contribution 
(proportional to the component of $M$ parallel to $c$) in the low-field regime caused by 
rotation of the Mn moment from planar (lying in the $ab$ plane) to uniaxial $(|| c)$.

Figure~\ref{fig5}(a) shows the procedures for extracting $\rho^{\rm A}_{xy}$ from the $\rho_{xy}$ data in the low and high temperature regimes;
Fig.~\ref{fig5}(b) shows the resulting values. As mentioned above, the anomalous Hall component associated with Ho
fluctuations close to the metamagnetic transition (red points) is largest at around $30-40$~K,
whereas that corresponding to rotation of the Mn moment from planar to uniaxial (blue points) grows with increasing temperature,
tracking the steep, low-field increase in magnetization~\cite{r23}.
These distinct behaviors suggests that the Mn and Ho spin textures influence the electrical conductivity differently.  

Finally, taking the high-field gradient of the 240~K Hall data [Fig~\ref{fig3}(d)] to represent the ordinary Hall contribution,
we use the Drude approximation $R_{\rm H} = 1/pe$~\cite{r37} to obtain an order-of-magnitude estimate for the
net hole density of $p \approx 7 \times 10^{27}~{\rm m}^{-3}$.

\begin{figure}
\begin{center}
\includegraphics[width=0.99\linewidth]{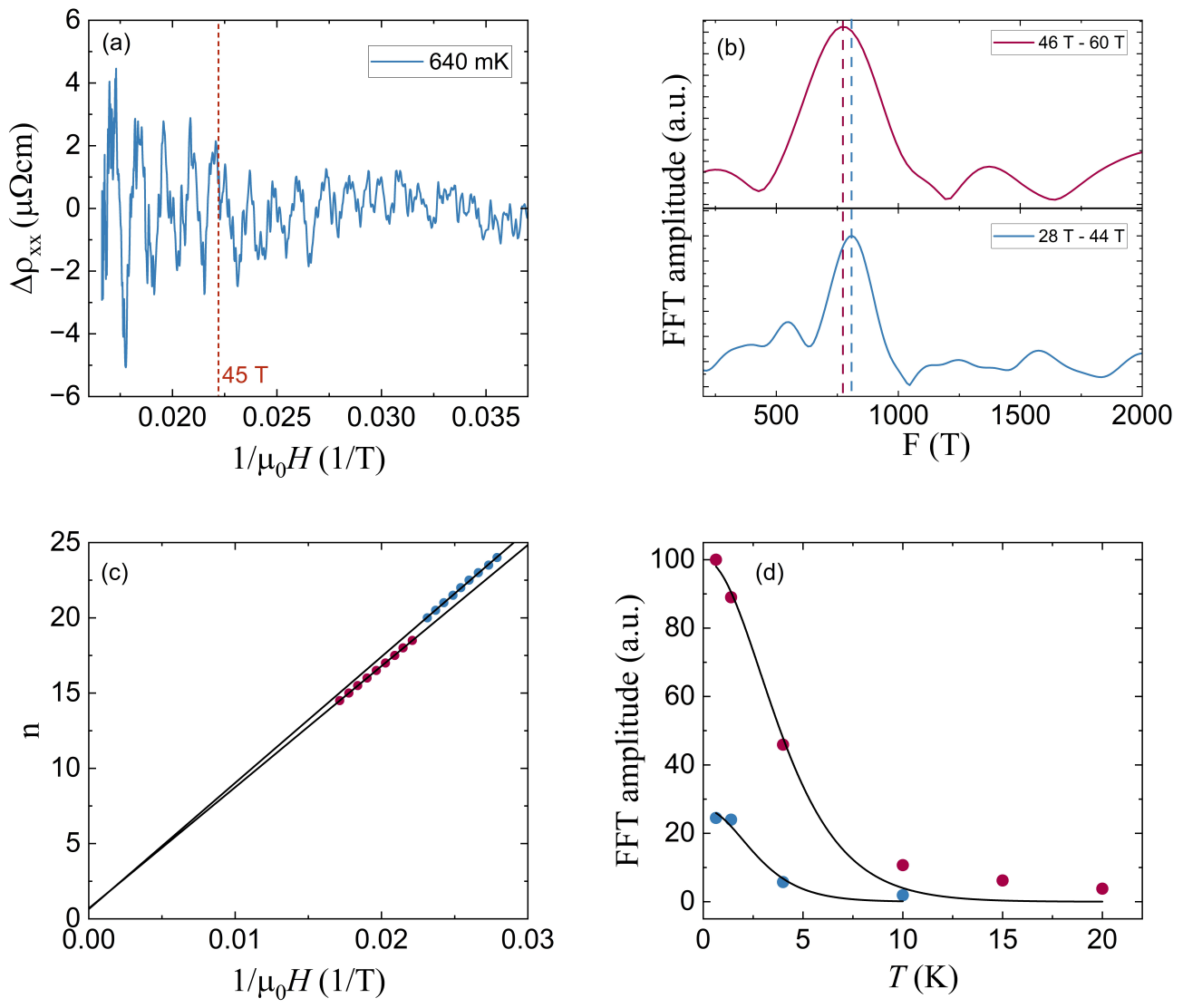}
\vspace{+4mm}
\caption{(a)~SdH oscillations shown as $\Delta \rho_{xx}$ versus $1/\mu_0 H$ for $T=640$~mK;
$\Delta \rho_{xx}$ is the oscillatory component of $\rho_{xx}$ obtained by subtracting the background magnetoresistance 
(fitted by a polynomial) from raw data.
(b)~Fourier spectra of the data in (a) taken using a $46-60$~T window (red) and 
a $26-44$~T window (blue), {\it i.e.,} above and below the metamagneetic transition, respectively.
(c)~Oscillation index plotted versus $1/H$ positions; red points are
from fields above the metamagnetic transition and blue are below it.
(d)~Temperature dependence of the SdH Fourier amplitudes at fields above (red)
and below (blue) the metamagnetic transition. The field windows used for the Fourier transforms are the same as those in (b).}
\vspace{-5mm}
\label{fig6}
\end{center}
\end{figure}

\subsection{Quantum oscillations}
We now consider the influence of the metamagnetic transition at 45~T on the SdH oscillation spectrum, which provides another measure of how the transition 
affects the underlying band structure and Fermi surface. 
The SdH quantum oscillations as a function of inverse applied magnetic field, $1/\mu_0 H$, are plotted in Fig.~\ref{fig6}(a). 
To assess the effect of the metamagnetic transition, Fourier analysis was performed over the field ranges $28~\text{--}~44$~T and $46~\text{--}~60$~T, 
obtaining single, dominant frequencies $F_{<45} = 836\pm3$~T and $F_{>45} = 801\pm8$~T, respectively,
 reflecting a 35~T shift in frequency at the transition [Fig.~\ref{fig6}(b)]. 
 Applying the Onsager relation~\cite{r38}, $F=\frac{\hbar S}{2 \pi e}$, where $S$ is the Fermi-surface cross-sectional area 
in the plane normal to $H$, suggests that $S$ changes from $6.40 \times 10^{18}~{\rm m^{-2}}$  to $6.16 \times 10^{18}~{\rm m^{-2}}$.
Such a shift in the oscillation frequency may in principle result from a jump in the internal field felt by the orbiting electrons due to the step in magnetization,
a change in carrier density (Fermi-surface volume) and/or a discontinuous shift in Berry phase reflecting an underlying alteration of the band topology. 
To distinguish between these possibilities, a Landau fan diagram was constructed by assigning $\rho_{xx}$ maxima
to integer $n$ and minima to $n+\frac{1}{2}$ in Fig.~\ref{fig6}(c). 
Indexing the Landau levels in this manner clearly reveals the frequency discontinuity at 45~T suggested by the Fourier analysis [see Fig.~\ref{fig6}(b)]. 
The straight-line fits [Fig.~\ref{fig6}(c)] show that the oscillations on both sides of the transition are accurately periodic in $1/H$, 
suggesting that changes in internal field due to the step in $M$ play only a very minor role (if any) in determining the position of the extrema. 
Therefore, the data were fitted using the Lifshitz–Onsager quantization criterion~\cite{r38}, $n= \frac{F}{\mu_0 H}+\beta$, giving $F = 838\pm9~$T and $\beta = 0.69\pm0.03$ below 45~T, 
and $F=808\pm3~$T and $\beta =0.65\pm0.06$ above 45~T; both derived frequencies are in good agreement with the values from Fourier analysis shown in Fig~\ref{fig6}(b). 
The change in slope (frequency) at 45 T without significant change of intercept, $\beta$, is consistent with a Fermi-surface contraction. This is supported by the program used to generate the Fourier transforms in Fig.~\ref{fig6}(a), which also calculates the phase of the oscillations; within experimental error, no phase change is detected at the metamagnetic transition.

Given that the effective mass derived from the temperature dependence of the quantum oscillations (see below) is $m^*\approx 2.0m_{\rm e}$, great caution must be exercised in attributing the intercept $\beta$ [Fig.~\ref{fig6}(c)] solely to a Berry phase.
The relevant term to consider from the Lifshitz-Kosevich formula, which describes the amplitude of the fundamental frequency of the quantum oscillations~\cite{r38}, is 
$\cos[\frac{1}{2} \pi g (m^*/m_{\rm e}]$; here, $g$ is the effective g-factor. When $1 < g (m^*/m_{\rm e}) <3$, this term returns a negative value, corresponding to an apparent phase change of $\pi$ in the oscillations. However, as $m^/m_{\rm e}\approx 2$, this would require a rather low value of $g$, considering the likely dominant role of Mn $d$-orbitals and Sn $s$ and $p$-orbitals in constructing the bandstructure. Instead, it is reasonable~\cite{r38} to propose that $g \approx 2$, yielding $\cos[\frac{1}{2} \pi g (m^*/m_{\rm e}]\approx 1$ ({\it i.e.,} a phase change of 0).
Nevertheless, in the current absence of bandstructure calculations for HoMn$_6$Sn$_6$, the Berry phase analysis that follows must be considered at best tentative.

In the absence of other effects, the parameter $\beta$ is related to the Berry phase $\phi_{\beta}$ through the relationship $\beta = \frac{\phi_{\beta}}{2\pi}\pm \delta$, where $\delta$ depends on 
the dimensionality of the Fermi surface and curvature of the extremal orbit. For a 3D band, $\delta = -\frac{1}{8}$ for a maximum (minimum) extremal orbit on an 
electron (hole) Fermi-surface pocket, and $+\frac{1}{8}$  for a minimum (maximum) extremal orbit on an electron (hole) pocket. 
From the positive sign of the Hall effect, it is probable, but not certain, that the observed oscillations originate from a hole band.
Because a Fermi-surface calculation that accounts for both the magnetic structure and the high magnetic field is not available, we considered all possible values of the Berry phase $\phi_\beta$ consistent with the measured $\beta$, as listed in Table~\ref{tbl:berry}. All of these values are nontrivial, and some of them, namely $\phi_\beta = 1.05\pi$ and $1.13\pi$, are close to $\pi$, suggesting a Dirac state.

\begin{table}[htbp]
\caption{Tentative possible Berry phases $\phi_\beta = 2\pi(\beta-\delta)$ for $\beta = 0.69$ and $0.65$ in a 3D band.}
\begin{tabular}{|c|c|c|c|}
\hline\hline
$\beta$ & $\delta$ & Orbit (electron / hole) & $\phi_\beta$ \\
\hline
0.69 & $-1/8$ & max / min & $1.63\pi$ \\
\hline
0.69 & $+1/8$ & min / max & $1.13\pi$ \\
\hline
0.65 & $-1/8$ & max / min & $1.55\pi$ \\
\hline
0.65 & $+1/8$ & min / max & $1.05\pi$ \\
\hline\hline
\end{tabular}
\label{tbl:berry}
\end{table}

Figure~\ref{fig6}(d) shows that the amplitude $A$ of the quantum oscillations decreases with increasing temperature. The corresponding cyclotron effective masses $m^*$ can be 
derived from these data using the component of the Lifshitz-Kosevich formula~\cite{r37,r38} that describes the thermal damping of the oscillations:
\begin{equation}
A(T) \propto \frac{\chi}{\sinh \chi},
\end{equation}
where, in SI units, $\chi = 14.69 m^*T/B_{\rm m}$, and $B_{\rm m} = 2/(B_1^{-1} + B_2^{-1})$ is the average inverse magnetic flux density from $B_1$ to $B_2$, the range over which 
the Fourier analysis was performed. (We assume that $B\approx \mu_0 H$ in the relatively high feld range of our data~\cite{r38}.)  Fitting the temperature dependence of the Fourier amplitudes over the field shown in Fig.~\ref{fig6}(b) 
yields $m^*/m_{\rm e} = 2.01 \pm 0.08$ for $\mu_0H > 45$~T and $m^*/m_{\rm e} =1.9\pm 0.3$ for $\mu_0 H < 45~$T, where $m_{\rm e}$ is the rest mass of a free electron.

Note that the Fermi-surface pocket measured by the SdH oscillations corresponds to only a small fraction of the carriers present.
Using a spherical Fermi-surface model to give an order-of-magnitude estimate, the measured cross-sections ranging from $6.16 \times 10^{18}~{\rm m^{-2}}$ 
to $6.4 \times 10^{18}~{\rm m^{-2}}$ correspond to a carrier density of just under $10^{26}~{\rm m^{-3}}$, a factor $\approx 70$ times smaller
than the effective hole density estimated from the Hall effect. There are very likely many other Fermi-surface pockets present; 
due to heavy effective masses, large scattering rates and/or geometries unconducive to extremal orbits, they are not observed in our data.

Finally, we emphasize once again that the metamagnetic transition does not seem to cause a significant change in the bandstructure, as evidenced by $\rho_{xy}$ [Fig.~\ref{fig4}(a)], the quantum oscillations [Fig.~\ref{fig6}(a-c)] and $\sigma_{xy}$ [Fig.~\ref{fig8}, shown in the Appendix].

\section{Discussion and Conclusions}
The measurements of HoMn$_6$Sn$_6$ in the current work were rendered possible by significant advances in two technological areas: (i)~the use of pulsed magnetic fields of up to 60~T, 
which, in conjunction with temperatures spanning 0.6~K to 240~K, allowed significant high-field features of the field-temperature phase diagram to be studied 
in detail using magnetometry for the first time; and (ii)~the successful FIB fabrication of strain-relieved devices that permitted sensitive transport studies 
to be carried out over the same ranges of temperature and magnetic field.
 
Perhaps the most significant discovery is that of a metamagnetic phase transition that shows a critical end-point at 45~T and 90~K, previously predicted by mean-field calculations~\cite{r26}. 
As far as we are aware, this is the only critical end-point that occurs within the $R$Mn$_6$Sn$_6$ kagome metals, and its temperature and field 
far exceed those of analogous phenomena in other materials. Whilst the magnetic field for the critical end-point predicted by the 
mean-field calculations is in good agreement with our data, the experimental temperature is somewhat higher that given by the calculations (50~K), perhaps due to Ho moment fluctuations.

As the critical end-point is approached from lower temperatures, the differential susceptibility ${\rm d}M/{\rm d}H$ at the phase boundary exhibits 
unusual temperature scaling, previously associated with magnetic fluctuations close to a quantum-critical point~\cite{r23,r24,r25}. 
Qualitative comparisons with other correlated-electron systems exhibiting analogous phenomena~\cite{r21,r35}
add further weight to our suggestion of the importance of magnetic fluctuations, as does the observation of a non-Fermi-liquid temperature dependence of the resistivity at
fields below the metamagnetic transition. Fermi-liquid behavior of the resistivity is recovered above the phase boundary associated with the critical end-point. 

At low temperatures, the metamagnetic transition that terminates at the critical end-point is accompanied by a small change in Fermi-surface configuration, shown by a
fall in the frequency of the Shubnikov-de Haas oscillations by roughly 4\%.
Neither phase of the oscillations nor the effective mass (which is close to $2.0m_{\rm e}$ in both phases) change significantly at the metamagnetic transition, suggesting that the latter has only a small effect on the bandstructure, a conclusion supported by the low-temperature field dependences of $\rho_{xy}$ and $\sigma_{xy}$.
We tentatively attribute the quantum-oscillation phase to nontrivial topology; a firmer conclusion must await detailed bandstructure calculations.

Note that the SdH oscillations only correspond to about $1/70$ of the total carrier density measured via Hall effect, which also indicates that the majority carriers are holes. 
The corresponding anomalous Hall effect does not correlate simply with the measured magnetization, showing that the magnetization (and its fluctuations) 
provided by the Ho and Mn sublattices independently affect the electrical conductivity in distinct ways.
 
In conclusion, we remark that previous measurements have explored a variety of possible magnetic phases and topological properties accessible in
 the RMn$_6$Sb$_6$ family by varying the rare-earth element R (R = Tb, Dy, Ho, Gd and Y)~\cite{r14,r21,r22,r24,r25}. By contrast, the experiments reported here, 
 underpinned by pulsed fields and novel sample processing, and supported by mean-field calculations, demonstrate that high magnetic fields provide another tuning 
 parameter to scan through different spin textures, this time within a single compound, potentially accessing new topological properties. Much exploration of parameter space remains.

 \section{Appendix}
 \subsection{Mean-field Hamiltonian and parameters}
Mean-field calculations of the magnetic phases in HoMn$_6$Sn$_6$ are based on the model Hamiltonian
\begin{equation}
\mathcal{H}=\mathcal{H}_{ex}+\mathcal{H}_{A}^{\rm Mn}+\mathcal{H}_{Z}^{\rm Mn}+\mathcal{H}_A^{\rm Ho}+ \mathcal{H}_{Z}^{\rm Ho}
\label{eqn:Hamiltonian}
\end{equation}
Here, $\mathcal{H}_{ex}$ are isotropic exchange interactions and $\mathcal{H}_{A}^{i}$ and $\mathcal{H}_{Z}^{i}$ are the single-ion anisotropy and Zeeman energies on site $i$, respectively. 

For a Ho site, these local terms are given by
\begin{equation}
\mathcal{H}_A^{\rm Ho} + \mathcal{H}_Z^{\rm Ho}=\sum_{lm}B_l^m\mathcal{O}_l^m({\bf J}_i) -g_J\mu_B{\bf J}_i\cdot {\mu_0\bf H} 
\label{eqn:HAR}
\end{equation}
where $B_l^m$ are the CEF parameters, $\mathcal{O}_l^m$ are the Stevens operators defined in terms of the total angular momentum ${\bf J}={\bf S}/(g_R-1)$.  For Ho,
 $J=8$, $S=2$ and the Land\'{e} $g$-factor is $g_J=5/4$.  For the $D_{6h}$ point symmetry of the $R$ ion, only  $B_2^0$, $B_4^0$, $B_6^0$, and $B_6^6$ are nonzero.  

For the Mn site, the local terms are
\begin{equation}
    \mathcal{H}_A^{\rm Mn} + \mathcal{H}_Z^{\rm Mn} = D^Ms_{z}^2 -g\mu_B{\bf s}\cdot \mu_0{\bf H}
\label{eqn:HZM}
\end{equation}
where $s=1$ and $g\approx 2.17$ is the Mn $g$-factor chosen to reproduce the measured magnetic moment.

The exchange Hamiltonian $\mathcal{H}_{ex}$ is given by
\begin{equation}
\mathcal{H}_{\rm ex}=\frac{1}{2}\sum_{i,j} \mathcal{J}^{MM}_{ij} \mathbf{s}_i \cdot \mathbf{s}_j + \frac{1}{2}\mathcal{J}^{MH} \sum_{i,j} \mathbf{s}_i \cdot \mathbf{S}_j
\label{eqn:Hex}
\end{equation}
where $\mathcal{J}^{MM}_{ij}$ represent various, and quite strong, intralayer and interlayer magnetic couplings between Mn 
spins (${\bf s}$). Here, $\mathcal{J}^{MH}>0$ is the AF coupling between neighboring Mn and Ho spins (${\bf S}$).

The values of the Hamiltonian parameters that are used and the details of the mean-field calculations are described in detail in \cite{r26}.  
Here, we repeat the parameter values (in meV); $B_2^0 = -0.00895$, $B_4^0 = 0.000314$, $B_6^0 = 0$, $B_6^6 = -1.01E$-5, $D^{M}=0.26$, 
$\mathcal{J}^{MM}_0  = -28.8$, $\mathcal{J}^{MM}_1  = -19.2$, $\mathcal{J}^{MM}_2  = -4.4$, $\mathcal{J}^{MM}_3  = 2.3$, $\mathcal{J}^{MH}  = 1.51$.
\begin{figure}
\begin{center}
\includegraphics[width=0.99\linewidth]{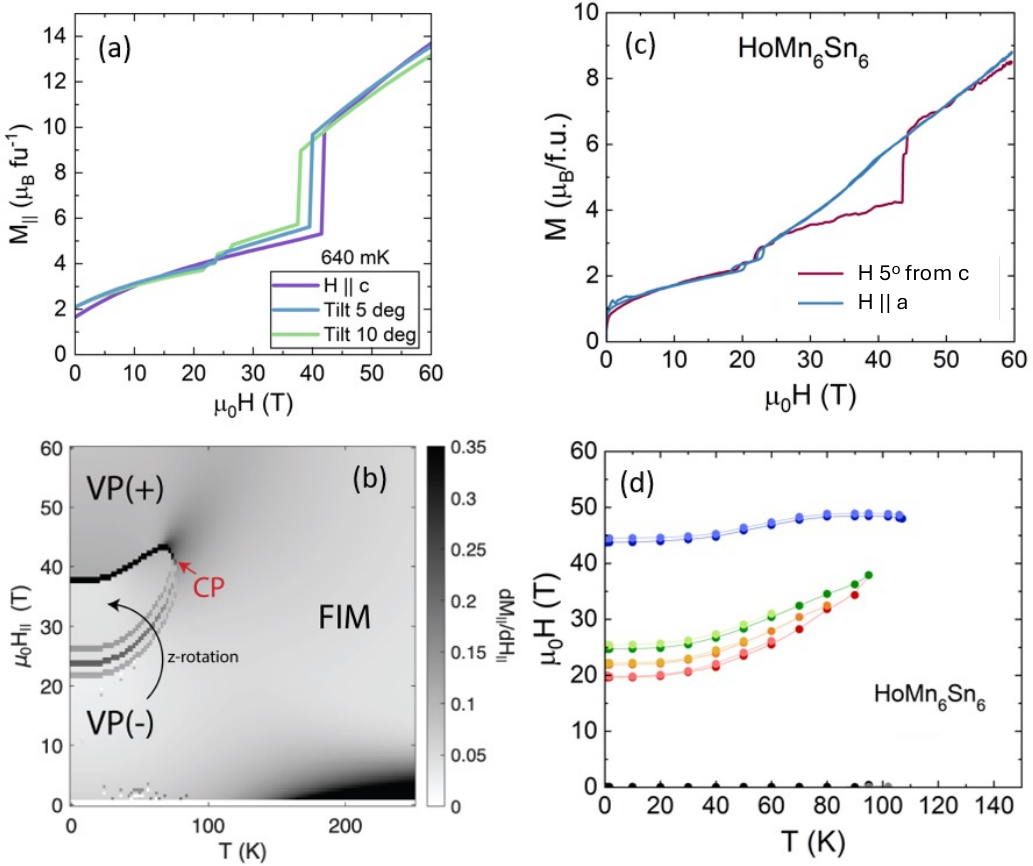}
\vspace{-4mm}
\caption{(a)~Simulated magnetization versus $H$ with small misalignement away from $c$ (see inset box for angles). (b)~theoretical contour plot predicted by
the mean-field model with $H$ tilted 10 degrees from the 
$c-$axis. (c)~Magnetization $M$ versus $H$ for $H||a$ and H tilted by about 5~degrees from $c$, recorded at 0.6~K. (d)~Experimental field versus temperature phase diagram for 
the slightly tilted crystal in (c). }
\vspace{-5mm}
\label{fig7}
\end{center}
\end{figure}
\subsection{Sensitivity of the magnetization to field alignment}
The calculations predict small magnetization steps observed between 20 and 30~T arising from slight misalignment of the applied 
field away from the crystallographic $c$-axis.  Figure~\ref{fig7}(a) depicts $M$ versus $H$ simulated data (from the mean-field model) for two instances of such misalignment; Figure~\ref{fig7}(b) shows the thoeretical phase diagram for magnetization that is parallel to an applied field tilted by 
$10^{\circ}$ away from the $c$-axis.  The strong first-order transition line near 40~T is now preceded by three first-order transitions [compare to Fig.~\ref{fig1}(b)]. 
These transitions are caused by rotation of the canted magnetic structure around the $c$-axis, as described below.

In the VP-canted phase, domains exist where the uncompensated planar magnetization points towards minima 
in the sixfold planar magnetic anisotropy energy. These domains are degenerate when there is no planar component of the magnetic field ($H_{\perp}=0$).  
However, small tilting of the field will introduce a planar field component such that these domains are no longer degenerate.  
At low fields, the Mn sublattice magnetization projected into the basal plane is dominant ($M_{\perp}^{\rm Mn} > -M_{\perp}^{\rm Ho}$) and the domain with 
$ H_{\perp} \parallel M_{\perp}^{\rm Mn}$ minimizes the Zeeman energy.  
As the $z$-component of the field increases, Fig.~\ref{fig7}(a) shows that the Mn moment rotates towards the field, decreasing $M_{\perp}^{\rm Mn}$, while the 
Ho magnetization is pinned by its large anisotropy energy.  
Near $\mu_0 H_z \approx  23$~T, the basal plane magnetization becomes compensated ($M_{\perp}^{\rm Mn} \approx -M_{\perp}^{\rm Ho}$), and ultimately 
Ho becomes dominant. With $H_{\perp} \neq 0$ and $H_z$ above the compensation field, the canted structure will rotate around the $c$-axis in discrete steps \
due to the sixfold planar anisotropy.  Figure \ref{fig7}(c) shows the experimental magnetization curves for a sample with $H$ accurately parallel to $a$
and one which has been deliberately tilted away from $H||c$ by about 5~degrees. The appearance of the predicted extra transitions is clear. The experimental phase diagram for the tilted sample is given in Fig.~\ref{fig7}(d).

\subsection{Hall conductivity}
\begin{figure}
\begin{center}
\includegraphics[width=0.99\linewidth]{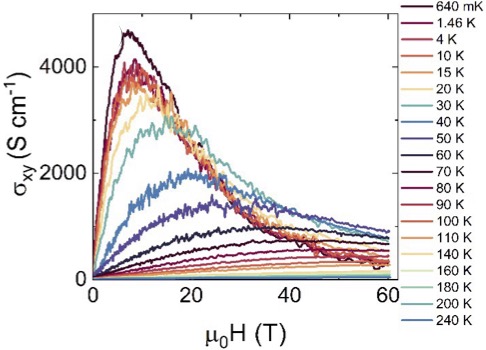}
\vspace{-4mm}
\caption{Experimental Hall conductivity $\sigma_{xy}$
versus field for several different temperatures.}
\vspace{-5mm}
\label{fig8}
\end{center}
\end{figure}

The Hall conductivity $\sigma_{xy}$ was calculated from 
simultaneously recorded experimental $\rho_{xx}$
and $\rho_{xy}$ data using the equation
\begin{equation}
    \sigma_{xy}=\frac{\rho_{xy}}{\rho_{xx}^2+\rho_{xy}^2}.
\end{equation}
The results are shown in Fig.~\ref{fig8}.
The large magnetization jump corresponding 
to the 45~T metamagnetic transition [Fig.~\ref{fig1}(b)] does not produce an effect visible in $\sigma_{xy}$.  This confirms that the jump seen in the Hall {\it resistivity}, $\rho_{xy}$ (see Fig.~\ref{fig4}), is a consequence of the anomalous Hall effect and not an indicator of changes in the Berry curvature across the transition.
This supports our interpretation of the observed change in the Shubnikov-de Haas oscillations at the metamagnetic transition [see discussion of Fig.~\ref{fig6}(c)].

\section{Acknowledgements}
The work was performed at the National High Magnetic Field Laboratory, which is supported by National Science Foundation Cooperative Agreement No.~DMR-2128556, the US Department of Energy and the State of Florida. JB, JS, TS, PC, RM and RJM are supported by the Center for the Advancement of Topological Semimetals, an Energy Frontier Research Center funded by the U.S.\ Department of Energy, Office of Science through the Ames National Laboratory under contract DE-AC02-07CH11358. JS acknowledges the DOE BES FWP ``Science of 100~T''.

\end{document}